\RequirePackage[l2tabu, orthodox]{nag}

\documentclass[
preprint,
 amsmath,amssymb,
 aps,
]{revtex4-1}

\usepackage{graphicx}
\usepackage{dcolumn}
\usepackage{bm}

\begin{document}

\preprint{APS}

\title{Dominance of tensor correlations in high-momentum nucleon pairs studied by ({\itshape p,pd}) reaction}
%
\author{S.~Terashima$^{1,2}$\email{tera@buaa.edu.cn}}
\author{L.~Yu$^{1}$}
\author{H.J.~Ong$^{3}$}
\author{I.~Tanihata$^{1,2,3}$}
\author{S.~Adachi$^{3}$}
\author{N.~Aoi$^{3}$}
\author{P.Y.~Chan$^{3}$}
\author{H.~Fujioka$^{4}$}
\author{M.~Fukuda$^{5}$}
\author{H.~Geissel$^{6,7}$}
\author{G.~Gey$^{3}$}
\author{J.~Golak$^{8}$}
\author{E.~Haettner$^{6,7}$}
\author{C.~Iwamoto$^{3}$}
\author{T.~Kawabata$^{4}$}
\author{H.~Kamada$^{9}$}
\author{X.Y.~Le$^{1,2}$}
\author{H.~Sakaguchi$^{3}$}
\author{A.~Sakaue$^{4}$}
\author{C.~Scheidenberger$^{6,7}$}
\author{R.~Skibi\'nski$^{8}$}
\author{B.H.~Sun$^{1,2,10}$}
\author{A.~Tamii$^{3}$}
\author{T.L.~Tang$^{3}$}
\author{D.T.~Tran$^{3,11}$}
\author{K.~Topolnicki$^{8}$}
\author{T.F.~Wang$^{1,2}$}
\author{Y.N.~Watanabe$^{12}$}
\author{H.~Weick$^{6}$}
\author{H.~Wita\l{}a$^{8}$}
\author{G.X.~Zhang$^{1,2}$}
\author{L.H.~Zhu$^{1,2,10}$}
\affiliation{$^1$School of Physics and Nuclear Energy Engineering, Beihang University, 
100191, Beijing, China}
\affiliation{$^2$International Research Center for Nuclei and Particles in Cosmos, 
Beihang University, 
100191, Beijing, China}
\affiliation{$^3$RCNP, Osaka University, 10-1 Mihogaoka, Ibaraki, Osaka 567-0047, Japan}
\affiliation{$^4$Department of Physics, Kyoto University, Kyoto 606-8502, Japan}
\affiliation{$^5$Osaka University, 1-5 Machikaneyama-cho, Toyonaka, Osaka 560-0043, Japan}
\affiliation{$^6$GSI Helmholtzzentrum f\"{u}r Schwerionenforschung GmbH, Planskstra$\ss$e 1, 64291
 Darmstadt, Germany}
\affiliation{$^7$Justus-Liebig-Universit\"{a}t Gie\ss en, Heinrich-Buff-Ring 16, 35392 Gie$\ss$en,
 Germany}
\affiliation{$^8$M. Smoluchowski Institute of Physics, Jagiellonian University, PL-30348 Krak\'{o}w, Poland}
\affiliation{$^{9}$Department of Physics, Faculty of Engineering, Kyushu Institute of Technology, Kitakyushu 804-8550, Japan}
\affiliation{$^{10}$Beijing Advanced Innovation Center for Big Data based Precision Medicine, Beihang University, 100083, Beijing, China}
\affiliation{$^{11}$Institute of Physics, Vietnam Academy of Science and Technology, Hanoi 100000, Vietnam}
\affiliation{$^{12}$Department of Physics, University of Tokyo, Tokyo 113-0033, Japan}

\date{\today}
\begin{abstract}

The isospin character of $p$-$n$ pairs at large relative momentum 
has been observed for the first time in the $^{16}$O ground state. 
A strong population of the $J$,$T$=1,0 state 
and a very weak population of the $J$,$T$=0,1 state 
were observed in neutron pick up domain of $^{16}$O($p$,$pd$) 
at 392 MeV. 
This strong isospin dependence at large momentum transfer is 
not reproduced by the distorted-wave 
impulse approximation calculations with known spectroscopic amplitudes.  
The results indicate the presence of high-momentum protons and neutrons 
induced by the tensor interactions in ground state of $^{16}$O.
\end{abstract}

\pacs{Valid PACS appear here}
\maketitle

High-momentum components in atomic nuclei are important for understanding the
roles of non-central nuclear interactions, such as the tensor interactions,
beyond the single particle motion characterized by the Fermi momentum. 
The importance of the tensor interactions, which act mainly between a proton
and a neutron in a nucleus, have been recognized from the binding energies of
light particles such as deuteron and alpha particle, 
and the presence of a large $D$-wave mixing in deuteron~\cite{Eri85}. 
Recent theoretical developments, particularly in {\itshape ab-initio}-type
calculations, 
enable treatment of 
high-momentum components directly including tensor interactions. 
An important feature in theoretical calculations~\cite{Sch07,Nef15} 
is a strong spin-isospin character of a pair of nucleons 
at large relative momentum in light nuclei. 

Extensive experimental studies 
via proton- and electron-induced reactions have been made at 
large momentum transfer~\cite{Tan03,Pia06,Shn07,Sub08,Kor14,Hen14,Due18}. 
These measurements show a dominance of $p$-$n$ over $p$-$p$ pairs 
at large relative momenta, which indicates the existence of 
short-range 
tensor correlations. 
Tensor correlations at large relative momenta could be 
extracted clearly by a spin-isospin correlated pair. 
Hence, identifying the spin-isospin of a pair is a key 
for distinguishing tensor 
and central interactions. 
The difference in cross sections between different spin-isospin states, 
denoted $S,T$=0,1 and $S,T$=1,0 in $p$-$n$ pairs, 
provides unique information on tensor correlations. 
The $S,T$=1,0 channel exhibits the strongest effect due to the 
tensor interaction, which may play a dominant role at higher momentum, 
especially around 2~fm$^{-1}$. 
In contrast, 
the $S,T$=0,1 channel 
has a negligible contribution 
from tensor interactions, 
with the main contribution coming from the central forces.

\begin{figure}[tbh]
\centering
\includegraphics[scale=.40]{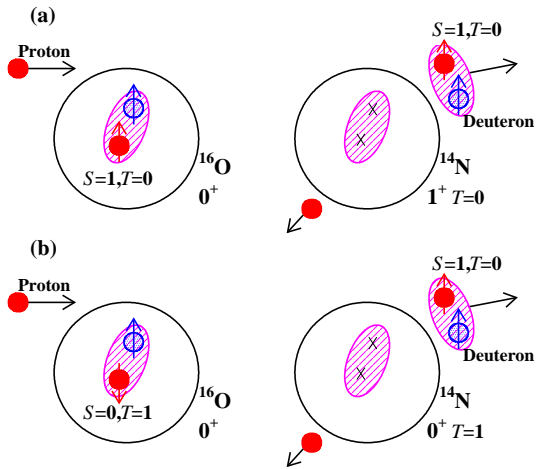}
\caption{Schematic view of neutron pick-up reaction with coincidence with a proton 
assuming (a) a $S,T$=1,0 correlated pair and 
(b) a $S,T$=0,1 correlated pair 
at the initial states in the ({\itshape p,pd}) reaction.}
\label{fig0}
\end{figure}

Recently, the ($p$,$d$) reaction at intermediate energies 
has been found to exhibit an effect of 
tensor interactions from the energy dependence of the cross sections 
populating specific excited states in $^{15}$O~\cite{Ong13}. 
An unexpected relative enhancement 
of the neutron pickup cross section 
feeding a positive parity state 
at large momentum transfers was detected, implying possible sensitivity 
of the ($p$,$d$) reaction 
to high-momentum components with tensor interactions in nuclei. 
A coincidence measurement ($p$,$Nd$) of a nucleon $N$ ($p$ or $n$) 
associated with a deuteron emitted at a small angle 
has strong sensitivity to correlated pairs of nucleons whose relative 
momentum is large. 
In the ({\itshape p,Nd}) reaction, we can distinguish the spin-isospin of the nucleon pairs $p$-$n$, and $n$-$n$ 
by measuring the specific final state of the residue with good energy resolution. 
Figure~\ref{fig0} shows a schematic view of the expected processes in 
the ({\itshape p,pd}) reaction. 
The pick-up mechanism of a neutron dominates when a scattered deuteron 
is observed at small angles.  If a reaction occurs with an $S$,$T$=1,0 pair, 
as shown in Fig.~\ref{fig0}(a), 
both nucleons are removed and thus the final state of 
the residual 
should have $T$=0.  
If instead the $p$-$n$ pair has $S$,$T$=0,1, the final state should be $T$=1, 
as shown in Fig.~\ref{fig0}(b). 

 ($p$,$pd$) reactions with forward deuteron detection have been reported 
for $^{6}$Li and $^{12}$C with 670-MeV incident protons~\cite{Alb80,Ero81}. 
However, the energy resolutions were 
insufficient to 
resolve the final states of the residual nuclei. 
Several studies on $^{12}$C and  $^{16}$O 
with low-energy protons have been reported~\cite{Des85,Gro77,Sam82}. 
Although the studies had sufficient resolution 
to resolve the individual states, 
the 
limited statistics and low incident energy for the quasi-free scattering 
makes it difficult to discuss the spin-isospin dependence 
in the ($p$,$pd$) reactions 
at large momentum transfer. 

In this paper, experimental results are presented
for high-momentum components 
observed in the ($p$,$pd$) reaction 
at high energy and at small deuteron scattering angle. 
The spin-isospin of the final states have been identified 
with high statistics and moderate energy resolution. 
The dominance of the $S$,$T$=1,0 channel in the cross section 
is observed 
for the first time in low-energy excited states of nuclei.
Another channel in ($p$,$nd$) was measured at the same time 
but the results will be described elsewhere. 

\begin{figure}[htb]
\centering
\includegraphics[scale=.4]{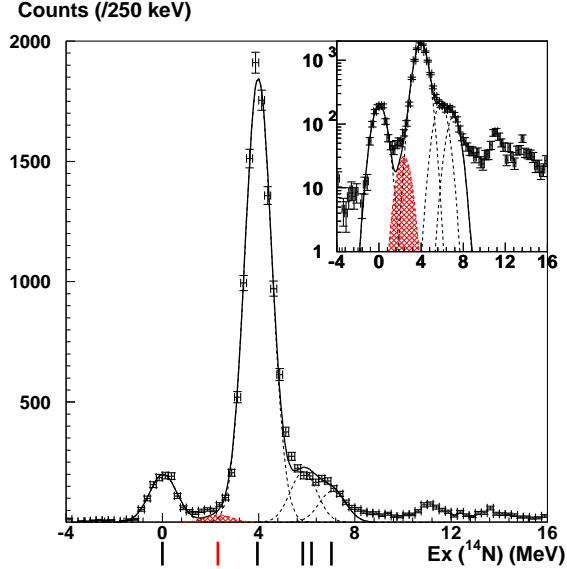}
\caption{The excitation energy spectrum of $^{16}$O({\itshape p,pd}) 
for $\theta_d$=8.6$^\circ$/$\theta_p$=138.4$^\circ$ 
with the total and individual fitting results 
shown by the solid and dashed lines, respectively.
}
\label{fig2}
\end{figure}

The experiment was performed at the West-South (WS) course of 
the Research Center for Nuclear Physics (RCNP) cyclotron facility 
using the newly constructed GRAF (Grand-RAiden Forward mode) 
beam line~\cite{GRAF}. 
Protons were accelerated to 392 MeV by the ring cyclotron and 
achromatically transported to the target 
in a scattering chamber. The beam spot was less than 1 mm in diameter. 
We used a windowless and self-supporting ice-sheet target~\cite{Kaw01} 
of thickness 56.2(4) mg/cm$^2$. 
Scattered deuterons were momentum analyzed by the high-resolution spectrometer 
Grand Raiden~\cite{Fuj99} equipped with 
two drift chambers and a pair of plastic scintillators on the focal plane. 
A typical intensity of incident beam was around a 20-nA. 
An excitation energy resolution of 260 keV full width at half maximum~(FWHM) 
in the ($p$,$d$) reaction was achieved, 
in which the limiting factors were stochastic energy loss in the target 
and the energy spread of the incident beam. 
The energy of the deuterons was calibrated over the whole acceptance range of the spectrometer 
using several transitions to discrete states in $^{15}$O. 
The accuracy of the scale was better than 50 keV. 
The coincidence detector array for protons 
consists of two 3-mm-thick ($\Delta E$) plastic scintillators 
and four horizontally 
segmented 60-mm-thick ($E$) and 240$\times$60-mm$^2$ plastic scintillator blocks.
The array covered a wide angular range corresponding to 240 mrad and 
90 mrad on the horizontal and vertical axes, respectively. 
The array was placed outside the scattering chamber through a thin window 
at backward angles 
to cover the zero recoil momenta of $^{14}$N in the $^{16}$O($p$,$pd$)$^{14}$N reaction. 
The excitation energies were determined from the momentum vectors of the detected deuterons 
and protons, where a coplanar geometry between deuterons and protons was assumed. 
The typical excitation energy spectrum of $^{14}$N is presented in Fig.~\ref{fig2}. 
An achieved energy resolution in ($p$,$pd$) reaction 1.6 MeV FWHM was dominated by the 
energy resolution of scintillators for protons. 

The ($p$,$d$) reaction at small angles 
is dominated by the pick up of high-momentum neutrons of correlated $p$-$n$ pairs. 
The correlated protons are 
emitted with a higher momentum at a backward angle, 
just like protons emitted in a backward angle in $p$+$d$ 
elastic scattering. 
The 2.31-MeV state in $^{14}$N is $J,T$=0,1
and the 3.95-MeV state is $J,T$=1,0;
and thus they are suitable for the present study. 
It should be noted that 
$J^\pi$ = 1$^+$ in the residue allows both $L$=0 and $L$=2 transitions, 
and the dominant transitions to the ground state and the second excited state 
at 3.95 MeV are 
considered to be $L$=2 and $L$=0, respectively, based on a theoretical study
by Cohen and Kurath~\cite{Coh70}. Thus, the ground state 1$^+$ with $L$=2 
does not satisfy the criteria used for the later analysis. 
Therefore, we focus only on the first and second excited states in this paper. 

Both states that can be described by a transferred angular momentum $L$=0 
in the $^{16}$O({\itshape p,pd}) reaction were 
observed with similar amplitudes in experiments with
75-MeV protons~\cite{Gro77}. 
In the present experiment the yield of the first excited state of 2.31~MeV 
is much lower than that of the 3.95-MeV state. 
The solid curves in Fig.~\ref{fig2} represent 
the spectra fittings assuming Gaussian line shapes with the same width. 
Here $E_x$ = 0.00, 2.31, 3.95, 5.83, 6.20, and 7.03 MeV 
below the proton separation energy 7.30 MeV in $^{14}$N, 
indicated by the vertical lines at the bottom of Fig.~\ref{fig2}, 
were assumed. 
The peak at 2.31 MeV is shown by the red hatched area, 
the contribution from 
the other states are shown by the dashed line, 
and the fitted curve including all the states is shown by the solid curve. 
The observed reduction of the $S,T$=0,1 state at 2.31 MeV 
is qualitatively consistent with the expected momentum dependence 
between $S,T$=1,0 and $S,T$=0,1, 
when the tensor interactions play an important role. 
We study this ratio in a more quantitative way by 
the distorted-wave impulse approximations (DWIA) in the following.

\begin{figure}[htb]
\centering
\includegraphics[scale=.48]{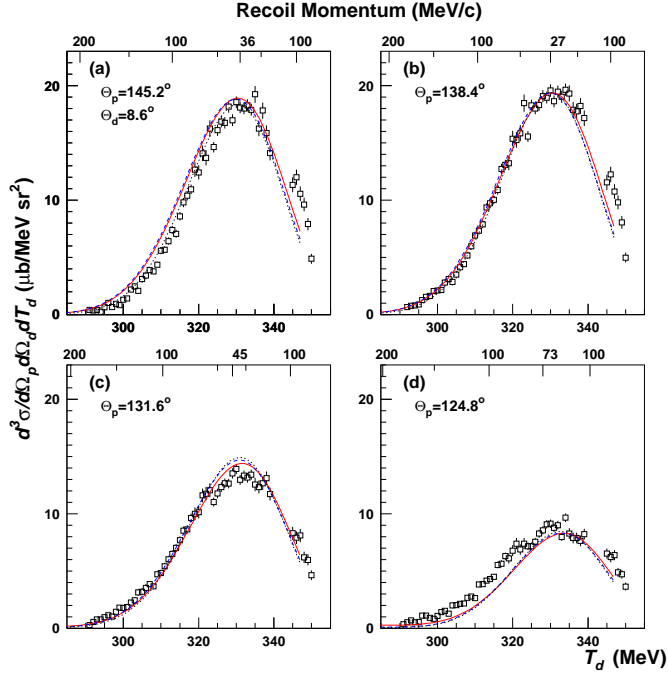}
\caption{Energy sharing spectra of deuterons around 4 MeV 
in the $^{16}$O({\itshape p,pd}) reaction at 
four different proton angles (a),(b),(c),(d). 
The upper abscissas provide scales for the averaged recoil momentum. 
The curves 
represent the DWIA calculations 
with the three different prescriptions, final energy (red-solid), 
initial energy (blue-dashed), and momentum-final energy (black-dotted).} 
\label{fig3}
\end{figure}

The experimental results were interpreted in terms of factorized amplitude 
DWIA using the code {\small THREEDEE}~\cite{Cha77}. 
The three-body triple differential cross section for $^{16}$O({\itshape p,pd}) 
is given as
\begin{equation}
\frac{d^3\sigma}{d\Omega_p d\Omega_d dT_d}
=S_d F_k\frac{d\sigma}{d\Omega}_{p+d}\sum_{\Lambda L} |T^\Lambda_L|^2
\label{eq1}
\end{equation}
for the particular cases 
of $S$=0 or $S$=$\frac{1}{2}$ and $L$=0~\cite{Gro77,Cha77}, 
where $F_k$ is a kinematic factor, 
and $S_d$ is the spectroscopic factor for deuteron in $^{16}$O. 
$\sum_{\Lambda L} |T^\Lambda_L|^2$ is the transition matrix, 
which 
is given by the overlapped function. 
The matrix contains the information of each quantum number 
of the deuteron, 
the relative angular momentum $L$ and its projection $\Lambda$.
The off-shell cross section of $p+d$ was approximated by the on-shell 
cross section with three conventional prescriptions called 
the initial energy, final energy, and 
momentum transfer-final energy prescriptions in the DWIA. 
We prepared the phenomenological optical potential of $p$+$d$ 
elastic scattering 
by introducing an $l$-dependent Majorana exchange term~\cite{Tho71,Vot74}, and 
fitting the data from the measurement with 392-MeV protons~\cite{Tam}.
Distorted waves of protons were calculated using the Schr\"{o}dinger equivalent 
reduction of the global potential from the Dirac phenomenology 
with the Darwin term~\cite{Coo93}. 
The optical potential deduced from the deuteron elastic scattering of 
$^{16}$O at 400 MeV~\cite{Ngu87} was used for the distorted waves 
of deuterons. 
The deuteron bound-state wave function generated using 
the Woods--Saxon potential had radius and diffuseness parameters 
of $r_0$ = 1.25~fm and $a_0$ = 0.65~fm, respectively. 
The well depth was adjusted to reproduce the separation energy. 
The Perey factor 
for a nonlocality correction~\cite{Per62} was applied at 0.54~fm 
for the deuteron scattering wave function and the bound state wave function. 

Figure \ref{fig3} shows the triple differential cross sections of 
$^{16}$O({\itshape p,pd})$^{14}$N  
as a function of deuteron energy for 
the region between 2 MeV and 6 MeV in excitation energy. 
Panels (a)--(d) show the cross sections obtained from four blocks of 
proton detectors covering different scattering angles, as indicated 
in the figure. 
The curves in Fig.~\ref{fig3} (a)--(d) are the results 
of the DWIA calculations for the 3.95-MeV state with $L$=0 transition. 
The upper abscissas in the figure provide scales for the averaged recoil momentum 
with angular acceptances of both the spectrometer and the detectors.  
One of the detectors whose data was shown in the panel (b) covered well 
the zero recoil part of the kinematics. 
The three different prescriptions, as described above, are presented 
by the three curves. 
The spectroscopic amplitude was adjusted by the one block at the peak nearest 
to the zero-recoil condition to reproduce the cross section 
They give almost the same results for all cases except for the absolute scales 
of the cross sections. 
As seen in Fig.~\ref{fig3}, 
the DWIA calculations describe the cross section very well, 
which confirms that the transition is dominantly $L$=0, 
and the mixing of $L$=2 is negligibly small, 
consistent with previous findings. 
This agreement also indicates that the background from 
a sequential decay with the two-body ($p$,$d$) transfer reaction 
was negligibly small in the present kinematical domain.

The state at 2.31~MeV in $T$=1 was also expected to be of $L$=0 transition. 
We therefore apply the quasi-free $^{16}$O($p$,$pd$)$^{14}$N 
reaction calculations with a similar procedure but replacing 
the off-shell cross sections from the $p$+$d$ elastic scattering 
with the more appropriate cross sections in the present DWIA.
In this case a correlated $p$-$n$ pair (denoted as $d^*$) has $S,T$=0,1, 
and the cross section of $d^*$($p$,$p$)$d$ was deduced 
from the detailed balance of the inverse reaction 
for the break-up $d$($p$,$pn$)$p$ reaction 
using the relevant kinematics. 
Unfortunately, no experimental data for the break-up reaction are available 
at energies higher than 26~MeV~\cite{Bru70}. Therefore, we deduced 
the cross section of a related reaction 
with the Faddeev calculation~\cite{Glo96} 
using the CD-Bonn interaction~\cite{Mar01} 
at 392~MeV. 
We followed the treatment in Refs.~\cite{Wee71,Bur72} 
and integrated the cross section with respect to the 
relative energy of $p$-$n$, which is a singlet state of deuteron $d^*$, 
up to 1 MeV. 
Similar calculations were performed for data at 75 MeV~\cite{Gro77} as references.
The differences in the energy dependence for the angular distribution of 
the elastic scattering $p$+$d$ and the $d^*$($p$,$p$)$d$ reaction 
introduce an energy dependence for the ratio between the two states. 
The ratio of the cross sections for the two reactions from around 37.5 degrees 
at 75 MeV to that at 392~MeV was estimated to be roughly a factor of 3 
in the present analysis. 

\begin{figure}
\centering
\includegraphics[scale=.45]{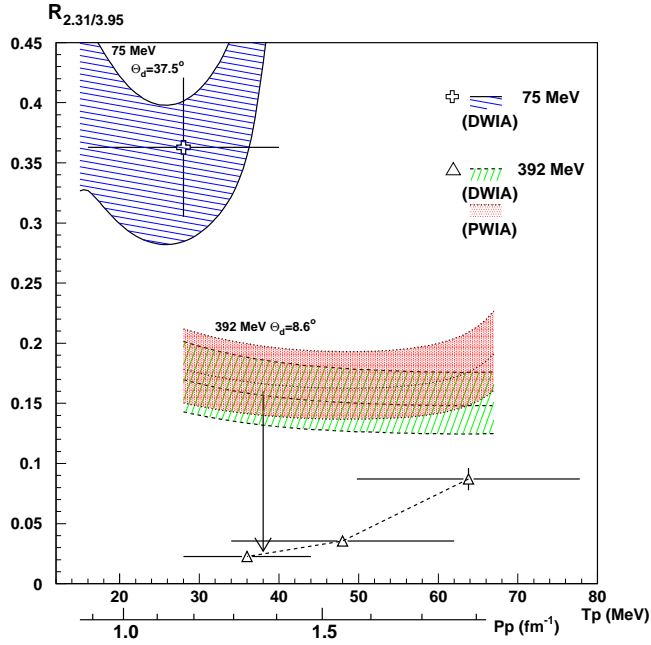}
\caption{Ratio of cross section for 2.31 MeV to 3.95 MeV in the sharing energy
spectra of protons in the $^{16}$O({\itshape p,pd}) reaction with different 
B$\rho$-settings. 
The vertical and horizontal bars of the data show the statistical error 
and detector coverage for each setting, 
respectively. See the text for details.}
\label{fig4}
\end{figure}

The triangular symbols in Fig.~\ref{fig4} show the ratio 
($R_{\rm 2.31/3.95}$) of 
the cross sections between the 2.31-MeV and 3.95-MeV states determined 
by the present experiment. The three data points are 
for different spectrometer settings 
and cover different proton momenta~($P_p$), 
where zero recoil momentum is indicated by the arrow. 
The ratio $R_{\rm 2.31/3.95}$ from the previous data at 75 MeV 
is denoted by the cross symbol. 
The results of DWIA calculations 
are shown by the blue hatched area. 
The ratio of the spectroscopic amplitudes between 2.31 MeV and 3.95 MeV 
was adjusted to reproduce the observed cross sections at 75 MeV.
The broadening of the red hatched area is due to the uncertainties 
of the data at 75 MeV~\cite{Gro77}. 
The results of the DWIA calculations at 392 MeV, applying 
the determined spectroscopic amplitude ratio 
are shown by the green hatched area. 
To see the sensitivity of the distorted waves in the calculation, 
the results of the plane-wave impulse approximation 
(PWIA) are also shown in the figure by the red dotted area. 
The ratio is not significantly different from that for DWIA, showing that the effect 
of distortion is small for the ratio $R_{\rm 2.31/3.95}$. 
The energies and angles of the emitted proton and deuteron are 
almost the same between the reactions to 
the first and second excited states, 
and therefore, the distortions essentially cancel out 
in the calculation of $R_{\rm 2.31/3.95}$. 
Due to the insensitivity to the distortion of the wave, 
we expect that the ratio $R_{\rm 2.31/3.95}$ 
closely reflects the ratio of the spectroscopic amplitude between the $S$,$T$=0,1 
and $S$,$T$=1,0 states.  The results of DWIA calculations indicate that 
the ratio $R_{\rm 2.31/3.95}$ is much smaller than that at 75 MeV 
due to the effect of tensor interactions. 
In other words, the apparent spectroscopic amplitude of 
$S$,$T$=0,1 is much smaller than that of $S$,$T$=1,0. 
It should be noted that the DWIA calculations include 
all nucleon--nucleon two-body interactions including tensor interactions.  
The present experimental values show the expected reduction of the 
ratio $R_{\rm 2.31/3.95}$ qualitatively 
but the ratio is as much as 5 times smaller than the DWIA prediction. 
The difference 
may also reflect a change in the tensor correlations in heavier nuclei. 
To understand 
this discrepancy, additional data 
on the $S$,$T$=1,0 and $S$,$T$=0,1 amplitudes 
as a function of the transferred momentum and nuclear mass 
are necessary. 
Further theoretical studies that include
more realistic structure information and reaction treatments 
are also anticipated. 

In summary, the cross sections of 
the $^{16}$O($p$,$pd$)$^{14}$N reactions were 
measured for 392-MeV incident protons with the coincidences between protons 
at backward angles 
and deuterons at forward angles, where the neutron pickup reaction mechanism is dominant. 
The first and second excited states in the residual 
$^{14}$N with $L$=0 transitions were compared with the DWIA calculations. 
A strong relative reduction of the first excited state cross section compared to 
that of the second excited state 
was observed, which is expected to be due to 
the tensor correlations. 
The DWIA calculations using the two-body interactions including tensor interactions 
qualitatively explained 
the reduction of the $S,T$=0,1 to $S,T$=1,0 ratio 
of the cross sections. 
However, 
The ratio of the experimental cross sections were 
overestimated by the calculations 
by as much as a factor of 5. 
Further experimental and theoretical studies 
are required to clarify the difference.

\begin{acknowledgments}
The authors thank the cyclotron crews at RCNP 
for their efforts to provide a clean and stable beam.
The support of the P.R.~China government and Beihang University 
under the Thousand Talent program is gratefully acknowledged. 
This work was partially supported by the National Natural Science 
Foundation of China under Contracts No.~11235002, No.~11375023, 
No.~11475014, and No.11575018, and by the National Key R\&D 
program of China (2016YFA0400504) 
and Hirose International Scholarship Foundation. 
The experiment was partly supported by a grant-in-aid program of the 
Japanese government under the contract numbers 23224008 and 20244030. 
\end{acknowledgments}


\bibliographystyle{apsrev-no}
\bibliography{tensor}

\begin{thebibliography}{32}
\expandafter\ifx\csname natexlab\endcsname\relax\def\natexlab#1{#1}\fi
\expandafter\ifx\csname bibnamefont\endcsname\relax
  \def\bibnamefont#1{#1}\fi
\expandafter\ifx\csname bibfnamefont\endcsname\relax
  \def\bibfnamefont#1{#1}\fi
\expandafter\ifx\csname citenamefont\endcsname\relax
  \def\citenamefont#1{#1}\fi
\expandafter\ifx\csname url\endcsname\relax
  \def\url#1{\texttt{#1}}\fi
\expandafter\ifx\csname urlprefix\endcsname\relax\def\urlprefix{URL }\fi
\providecommand{\bibinfo}[2]{#2}
\providecommand{\eprint}[2][]{\url{#2}}

\bibitem[{\citenamefont{Ericson and Rosa-Clot}(1985)}]{Eri85}
\bibinfo{author}{\bibfnamefont{T.~E.~O.} \bibnamefont{Ericson}}
  \bibnamefont{and}
  \bibinfo{author}{\bibfnamefont{M.}~\bibnamefont{Rosa-Clot}},
  \bibinfo{journal}{Annu. Rev. Nucl. Sci.} \textbf{\bibinfo{volume}{35}},
  \bibinfo{pages}{271} (\bibinfo{year}{1985}).

\bibitem[{\citenamefont{Schiavilla {\it et~al.}}(2007)\citenamefont{Schiavilla,
  Wiringa, Pieper, and Carlson}}]{Sch07}
\bibinfo{author}{\bibfnamefont{R.}~\bibnamefont{Schiavilla}},
  \bibinfo{author}{\bibfnamefont{R.~B.} \bibnamefont{Wiringa}},
  \bibinfo{author}{\bibfnamefont{S.~C.} \bibnamefont{Pieper}},
  \bibnamefont{and} \bibinfo{author}{\bibfnamefont{J.}~\bibnamefont{Carlson}},
  \bibinfo{journal}{Phys. Rev. Lett.} \textbf{\bibinfo{volume}{98}},
  \bibinfo{pages}{132501} (\bibinfo{year}{2007}).

\bibitem[{\citenamefont{Neff {\it et~al.}}(2015)\citenamefont{Neff, Feldmeier,
  and Horiuchi}}]{Nef15}
\bibinfo{author}{\bibfnamefont{T.}~\bibnamefont{Neff}},
  \bibinfo{author}{\bibfnamefont{H.}~\bibnamefont{Feldmeier}},
  \bibnamefont{and} \bibinfo{author}{\bibfnamefont{W.}~\bibnamefont{Horiuchi}},
  \bibinfo{journal}{Phys. Rev. C} \textbf{\bibinfo{volume}{92}},
  \bibinfo{pages}{024003} (\bibinfo{year}{2015}).

\bibitem[{\citenamefont{Tang {\it et~al.}}(2003)\citenamefont{Tang, Watson,
  Aclander, Alster, Asryan, Averichev, Barton, Baturin, Bukhtoyarova, Carroll
  {\it et~al.}}}]{Tan03}
\bibinfo{author}{\bibfnamefont{A.}~\bibnamefont{Tang}}, \bibnamefont{{\it
  et~al.}}, \bibinfo{journal}{Phys. Rev. Lett.} \textbf{\bibinfo{volume}{90}},
  \bibinfo{pages}{042301} (\bibinfo{year}{2003}).

\bibitem[{\citenamefont{Piasetzky {\it et~al.}}(2006)\citenamefont{Piasetzky,
  Sargsian, Frankfurt, Strikman, and Watson}}]{Pia06}
\bibinfo{author}{\bibfnamefont{E.}~\bibnamefont{Piasetzky}},
  \bibinfo{author}{\bibfnamefont{M.}~\bibnamefont{Sargsian}},
  \bibinfo{author}{\bibfnamefont{L.}~\bibnamefont{Frankfurt}},
  \bibinfo{author}{\bibfnamefont{M.}~\bibnamefont{Strikman}}, \bibnamefont{and}
  \bibinfo{author}{\bibfnamefont{J.~W.} \bibnamefont{Watson}},
  \bibinfo{journal}{Phys. Rev. Lett.} \textbf{\bibinfo{volume}{97}},
  \bibinfo{pages}{162504} (\bibinfo{year}{2006}).

\bibitem[{\citenamefont{Shneor {\it et~al.}}(2007)\citenamefont{Shneor,
  Monaghan, Subedi, Anderson, Aniol, Annand, Arrington, Benaoum, Benmokhtar,
  Bertin {\it et~al.}}}]{Shn07}
\bibinfo{author}{\bibfnamefont{R.}~\bibnamefont{Shneor}}, \bibnamefont{{\it
  et~al.}} (\bibinfo{collaboration}{Jefferson Lab Hall A Collaboration}),
  \bibinfo{journal}{Phys. Rev. Lett.} \textbf{\bibinfo{volume}{99}},
  \bibinfo{pages}{072501} (\bibinfo{year}{2007}).

\bibitem[{\citenamefont{Subedi {\it et~al.}}(2008)\citenamefont{Subedi, Shneor,
  Monaghan, Anderson, Aniol, Annand, Arrington, Benaoum, Benmokhtar, Boeglin
  {\it et~al.}}}]{Sub08}
\bibinfo{author}{\bibfnamefont{R.}~\bibnamefont{Subedi}}, \bibnamefont{{\it
  et~al.}}, \bibinfo{journal}{Science} \textbf{\bibinfo{volume}{320}},
  \bibinfo{pages}{1476} (\bibinfo{year}{2008}).

\bibitem[{\citenamefont{Korover {\it et~al.}}(2014)\citenamefont{Korover,
  Muangma, Hen, Shneor, Sulkosky, Kelleher, Gilad, Higinbotham, Piasetzky,
  Watson {\it et~al.}}}]{Kor14}
\bibinfo{author}{\bibfnamefont{I.}~\bibnamefont{Korover}}, \bibnamefont{{\it
  et~al.}} (\bibinfo{collaboration}{Jefferson Lab Hall A Collaboration}),
  \bibinfo{journal}{Phys. Rev. Lett.} \textbf{\bibinfo{volume}{113}},
  \bibinfo{pages}{022501} (\bibinfo{year}{2014}).

\bibitem[{\citenamefont{Hen {\it et~al.}}(2014)\citenamefont{Hen, Sargsian,
  Weinstein, Piasetzky, Hakobyan, Higinbotham, Braverman, Brooks, Gilad,
  Adhikari {\it et~al.}}}]{Hen14}
\bibinfo{author}{\bibfnamefont{O.}~\bibnamefont{Hen}}, \bibnamefont{{\it
  et~al.}} (\bibinfo{collaboration}{Jefferson Lab CLAS Collaboration}),
  \bibinfo{journal}{Science} \textbf{\bibinfo{volume}{346}},
  \bibinfo{pages}{614} (\bibinfo{year}{2014}).

\bibitem[{\citenamefont{Duer {\it et~al.}}(2018)\citenamefont{Duer, Hen,
  Piasetzky, Hakobyan, Weinstein, Braverman, Cohen, Higinbotham, Adhikari,
  Adhikari {\it et~al.}}}]{Due18}
\bibinfo{author}{\bibfnamefont{M.}~\bibnamefont{Duer}}, \bibnamefont{{\it
  et~al.}} (\bibinfo{collaboration}{Jefferson Lab CLAS Collaboration}),
  \bibinfo{journal}{Nature} \textbf{\bibinfo{volume}{560}},
  \bibinfo{pages}{617} (\bibinfo{year}{2018}).

\bibitem[{\citenamefont{Ong {\it et~al.}}(2013)\citenamefont{Ong, Tanihata,
  Tamii, Myo, Ogata, Fukuda, Hirota, Ikeda, Ishikawa, Kawabata {\it
  et~al.}}}]{Ong13}
\bibinfo{author}{\bibfnamefont{H.~J.} \bibnamefont{Ong}}, \bibnamefont{{\it
  et~al.}}, \bibinfo{journal}{Phys. Lett. B} \textbf{\bibinfo{volume}{725}},
  \bibinfo{pages}{277 } (\bibinfo{year}{2013}).

\bibitem[{\citenamefont{Albrecht {\it et~al.}}(1980)\citenamefont{Albrecht,
  Csatl{\'o}s, Er{\"o}, Fodor, Hernyes, Hongsung, Khomenko, Khovanskij, Koncz,
  Krumstein {\it et~al.}}}]{Alb80}
\bibinfo{author}{\bibfnamefont{D.}~\bibnamefont{Albrecht}}, \bibnamefont{{\it
  et~al.}}, \bibinfo{journal}{Nucl. Phys. A} \textbf{\bibinfo{volume}{338}},
  \bibinfo{pages}{477 } (\bibinfo{year}{1980}).

\bibitem[{\citenamefont{Er{\"o} {\it et~al.}}(1981)\citenamefont{Er{\"o},
  Fodor, Koncz, Seres, Csatl{\'o}s, Khomenko, Khovanskij, Krumstein, Merekov,
  and Petrukhin}}]{Ero81}
\bibinfo{author}{\bibfnamefont{J.}~\bibnamefont{Er{\"o}}}, \bibnamefont{{\it
  et~al.}}, \bibinfo{journal}{Nucl. Phys. A} \textbf{\bibinfo{volume}{372}},
  \bibinfo{pages}{317 } (\bibinfo{year}{1981}).

\bibitem[{\citenamefont{Descroix {\it et~al.}}(1985)\citenamefont{Descroix,
  Bedjidian, Grossiord, Guichard, Gusakow, Jacquin, Pizzi, and Bagieu}}]{Des85}
\bibinfo{author}{\bibfnamefont{E.}~\bibnamefont{Descroix}}, \bibnamefont{{\it
  et~al.}}, \bibinfo{journal}{Nucl. Phys. A} \textbf{\bibinfo{volume}{438}},
  \bibinfo{pages}{112 } (\bibinfo{year}{1985}).

\bibitem[{\citenamefont{Grossiord {\it et~al.}}(1977)\citenamefont{Grossiord,
  Bedjidian, Guichard, Gusakow, Pizzi, Delbar, Gr\'egoire, and Lega}}]{Gro77}
\bibinfo{author}{\bibfnamefont{J.~Y.} \bibnamefont{Grossiord}},
  \bibnamefont{{\it et~al.}}, \bibinfo{journal}{Phys. Rev. C}
  \textbf{\bibinfo{volume}{15}}, \bibinfo{pages}{843} (\bibinfo{year}{1977}).

\bibitem[{\citenamefont{Samanta {\it et~al.}}(1982)\citenamefont{Samanta,
  Chant, Roos, Nadasen, and Cowley}}]{Sam82}
\bibinfo{author}{\bibfnamefont{C.}~\bibnamefont{Samanta}},
  \bibinfo{author}{\bibfnamefont{N.~S.} \bibnamefont{Chant}},
  \bibinfo{author}{\bibfnamefont{P.~G.} \bibnamefont{Roos}},
  \bibinfo{author}{\bibfnamefont{A.}~\bibnamefont{Nadasen}}, \bibnamefont{and}
  \bibinfo{author}{\bibfnamefont{A.~A.} \bibnamefont{Cowley}},
  \bibinfo{journal}{Phys. Rev. C} \textbf{\bibinfo{volume}{26}},
  \bibinfo{pages}{1379} (\bibinfo{year}{1982}).

\bibitem[{\citenamefont{Iwamoto}()}]{GRAF}
\bibinfo{author}{\bibfnamefont{C.}~\bibnamefont{Iwamoto}},
  \bibinfo{note}{annual Report RCNP, 2014 (unpublished)}.

\bibitem[{\citenamefont{Kawabata {\it et~al.}}(2001)\citenamefont{Kawabata,
  Akimune, Fujimura, Fujita, Fujita, Fujiwara, Hara, Hatanaka, Hosono, Ishikawa
  {\it et~al.}}}]{Kaw01}
\bibinfo{author}{\bibfnamefont{T.}~\bibnamefont{Kawabata}}, \bibnamefont{{\it
  et~al.}}, \bibinfo{journal}{Nucl. Instrum. Meth. A}
  \textbf{\bibinfo{volume}{459}}, \bibinfo{pages}{171 } (\bibinfo{year}{2001}).

\bibitem[{\citenamefont{Fujiwara {\it et~al.}}(1999)\citenamefont{Fujiwara,
  Akimune, Daito, Fujimura, Fujita, Hatanaka, Ikegami, Katayama, Nagayama,
  Matsuoka {\it et~al.}}}]{Fuj99}
\bibinfo{author}{\bibfnamefont{M.}~\bibnamefont{Fujiwara}}, \bibnamefont{{\it
  et~al.}}, \bibinfo{journal}{Nucl. Instrum. Meth. A}
  \textbf{\bibinfo{volume}{422}}, \bibinfo{pages}{484 } (\bibinfo{year}{1999}).

\bibitem[{\citenamefont{Cohen and Kurath}(1970)}]{Coh70}
\bibinfo{author}{\bibfnamefont{S.}~\bibnamefont{Cohen}} \bibnamefont{and}
  \bibinfo{author}{\bibfnamefont{D.}~\bibnamefont{Kurath}},
  \bibinfo{journal}{Nucl. Phys. A} \textbf{\bibinfo{volume}{141}},
  \bibinfo{pages}{145 } (\bibinfo{year}{1970}).

\bibitem[{\citenamefont{Chant and Roos}(1977)}]{Cha77}
\bibinfo{author}{\bibfnamefont{N.~S.} \bibnamefont{Chant}} \bibnamefont{and}
  \bibinfo{author}{\bibfnamefont{P.~G.} \bibnamefont{Roos}},
  \bibinfo{journal}{Phys. Rev. C} \textbf{\bibinfo{volume}{15}},
  \bibinfo{pages}{57} (\bibinfo{year}{1977}).

\bibitem[{\citenamefont{Thompson and Tang}(1971)}]{Tho71}
\bibinfo{author}{\bibfnamefont{D.~R.} \bibnamefont{Thompson}} \bibnamefont{and}
  \bibinfo{author}{\bibfnamefont{Y.~C.} \bibnamefont{Tang}},
  \bibinfo{journal}{Phys. Rev. C} \textbf{\bibinfo{volume}{4}},
  \bibinfo{pages}{306} (\bibinfo{year}{1971}).

\bibitem[{\citenamefont{Votta {\it et~al.}}(1974)\citenamefont{Votta, Roos,
  Chant, and Woody}}]{Vot74}
\bibinfo{author}{\bibfnamefont{L.~G.} \bibnamefont{Votta}},
  \bibinfo{author}{\bibfnamefont{P.~G.} \bibnamefont{Roos}},
  \bibinfo{author}{\bibfnamefont{N.~S.} \bibnamefont{Chant}}, \bibnamefont{and}
  \bibinfo{author}{\bibfnamefont{R.}~\bibnamefont{Woody}},
  \bibinfo{journal}{Phys. Rev. C} \textbf{\bibinfo{volume}{10}},
  \bibinfo{pages}{520} (\bibinfo{year}{1974}).

\bibitem[{\citenamefont{Tamii {\it et~al.}}(2007)\citenamefont{Tamii, Dozono,
  Fujita, Hatanaka, Ihara, Kaneda, Kuboki, Maeda, Matsubara, Ohta {\it
  et~al.}}}]{Tam}
\bibinfo{author}{\bibfnamefont{A.}~\bibnamefont{Tamii}}, \bibnamefont{{\it
  et~al.}}, \bibinfo{organization}{AIP Conf. Proc. {\bf 915}, 765}
  (\bibinfo{year}{2007}).

\bibitem[{\citenamefont{Cooper {\it et~al.}}(1993)\citenamefont{Cooper, Hama,
  Clark, and Mercer}}]{Coo93}
\bibinfo{author}{\bibfnamefont{E.~D.} \bibnamefont{Cooper}},
  \bibinfo{author}{\bibfnamefont{S.}~\bibnamefont{Hama}},
  \bibinfo{author}{\bibfnamefont{B.~C.} \bibnamefont{Clark}}, \bibnamefont{and}
  \bibinfo{author}{\bibfnamefont{R.~L.} \bibnamefont{Mercer}},
  \bibinfo{journal}{Phys. Rev. C} \textbf{\bibinfo{volume}{47}},
  \bibinfo{pages}{297} (\bibinfo{year}{1993}).

\bibitem[{\citenamefont{van Sen {\it et~al.}}(1987)\citenamefont{van Sen,
  Yanlin, Arvieux, Gaillard, Bonin, Boudard, Bruge, Lugol, Hasegawa, Soga {\it
  et~al.}}}]{Ngu87}
\bibinfo{author}{\bibfnamefont{N.}~\bibnamefont{van Sen}}, \bibnamefont{{\it
  et~al.}}, \bibinfo{journal}{Nucl. Phys. A} \textbf{\bibinfo{volume}{464}},
  \bibinfo{pages}{717 } (\bibinfo{year}{1987}).

\bibitem[{\citenamefont{Perey and Buck}(1962)}]{Per62}
\bibinfo{author}{\bibfnamefont{F.}~\bibnamefont{Perey}} \bibnamefont{and}
  \bibinfo{author}{\bibfnamefont{B.}~\bibnamefont{Buck}},
  \bibinfo{journal}{Nucl. Phys.} \textbf{\bibinfo{volume}{32}},
  \bibinfo{pages}{353 } (\bibinfo{year}{1962}).

\bibitem[{\citenamefont{Br{\"u}ckmann {\it
  et~al.}}(1970)\citenamefont{Br{\"u}ckmann, Kluge, Matth{\"a}y, Sch{\"a}nzler,
  and Wick}}]{Bru70}
\bibinfo{author}{\bibfnamefont{H.}~\bibnamefont{Br{\"u}ckmann}},
  \bibinfo{author}{\bibfnamefont{W.}~\bibnamefont{Kluge}},
  \bibinfo{author}{\bibfnamefont{H.}~\bibnamefont{Matth{\"a}y}},
  \bibinfo{author}{\bibfnamefont{L.}~\bibnamefont{Sch{\"a}nzler}},
  \bibnamefont{and} \bibinfo{author}{\bibfnamefont{K.}~\bibnamefont{Wick}},
  \bibinfo{journal}{Nucl. Phys. A} \textbf{\bibinfo{volume}{157}},
  \bibinfo{pages}{209 } (\bibinfo{year}{1970}).

\bibitem[{\citenamefont{Gl\"{o}ckle {\it
  et~al.}}(1996)\citenamefont{Gl\"{o}ckle, Wita\l{}a, H\"{u}ber, Kamada, and
  Golak}}]{Glo96}
\bibinfo{author}{\bibfnamefont{W.}~\bibnamefont{Gl\"{o}ckle}},
  \bibinfo{author}{\bibfnamefont{H.}~\bibnamefont{Wita\l{}a}},
  \bibinfo{author}{\bibfnamefont{D.}~\bibnamefont{H\"{u}ber}},
  \bibinfo{author}{\bibfnamefont{H.}~\bibnamefont{Kamada}}, \bibnamefont{and}
  \bibinfo{author}{\bibfnamefont{J.}~\bibnamefont{Golak}},
  \bibinfo{journal}{Phys. Rep.} \textbf{\bibinfo{volume}{274}},
  \bibinfo{pages}{107 } (\bibinfo{year}{1996}).

\bibitem[{\citenamefont{Machleidt}(2001)}]{Mar01}
\bibinfo{author}{\bibfnamefont{R.}~\bibnamefont{Machleidt}},
  \bibinfo{journal}{Phys. Rev. C} \textbf{\bibinfo{volume}{63}},
  \bibinfo{pages}{024001} (\bibinfo{year}{2001}).

\bibitem[{\citenamefont{van~der Weerd {\it et~al.}}(1971)\citenamefont{van~der
  Weerd, Canada, Fink, and Cohen}}]{Wee71}
\bibinfo{author}{\bibfnamefont{J.~C.} \bibnamefont{van~der Weerd}},
  \bibinfo{author}{\bibfnamefont{T.~R.} \bibnamefont{Canada}},
  \bibinfo{author}{\bibfnamefont{C.~L.} \bibnamefont{Fink}}, \bibnamefont{and}
  \bibinfo{author}{\bibfnamefont{B.~L.} \bibnamefont{Cohen}},
  \bibinfo{journal}{Phys. Rev. C} \textbf{\bibinfo{volume}{3}},
  \bibinfo{pages}{66} (\bibinfo{year}{1971}).

\bibitem[{\citenamefont{Burq {\it et~al.}}(1972)\citenamefont{Burq, Cabrillat,
  Chemarin, Ille, and Nicolai}}]{Bur72}
\bibinfo{author}{\bibfnamefont{J.}~\bibnamefont{Burq}},
  \bibinfo{author}{\bibfnamefont{J.}~\bibnamefont{Cabrillat}},
  \bibinfo{author}{\bibfnamefont{M.}~\bibnamefont{Chemarin}},
  \bibinfo{author}{\bibfnamefont{B.}~\bibnamefont{Ille}}, \bibnamefont{and}
  \bibinfo{author}{\bibfnamefont{G.}~\bibnamefont{Nicolai}},
  \bibinfo{journal}{Nucl. Phys. A} \textbf{\bibinfo{volume}{179}},
  \bibinfo{pages}{371 } (\bibinfo{year}{1972}).

\end{thebibliography}

\end{document}